\documentclass[a4paper,amsmath,amssymb,amsfonts]{jpconf}
\usepackage{graphicx}

\begin{document}
\title{$f(G)$ modified gravity and the energy conditions}

\author{Nadiezhda M. Garc\'ia, Francisco S. N. Lobo, Jos\'e P. Mimoso}
\address{Centro de Astronomia e Astrof\'{\i}sica da Universidade de Lisboa,\\ Campo Grande, Ed. C8 1749-016 Lisboa, Portugal}
\ead{nadiezhda@cosmo.fis.fc.ul.pt, flobo@cii.fc.ul.pt, jpmimoso@cii.fc.ul.pt}

\author{Tiberiu Harko}
\address{Department of Physics and
Center for Theoretical and Computational Physics,\\ The University
of Hong Kong, Pok Fu Lam Road, Hong Kong}
\ead{harko@hkucc.hku.hk}

\begin{abstract}
Motivated by string/M-theory predictions that scalar field couplings with the Gauss-Bonnet invariant, $G$, are essential in the appearance of non-singular early time cosmologies, we discuss the viability of an interesting alternative gravitational theory, namely, modified Gauss-Bonnet gravity, and present the viability bounds arising from the energy conditions. In particular, we consider a specific realistic form of $f(G)$ analyzed in the literature that accounts for the late-time cosmic acceleration and that has been found to cure the finite-time future singularities present in the dark energy models, and further examine the respective viability of the specific $f(G)$ model imposed by the weak energy condition.
\end{abstract}

\section{Introduction}

A central theme in cosmology is the perplexing fact that the Universe is
undergoing an accelerating expansion \cite{expansion}. Several candidates, responsible for this expansion, have been proposed in the literature, in particular, dark energy models and modified gravity. In this context, a renaissance of modified theories of gravity, such as $f(R)$ gravity \cite{review}, has recently been verified in an attempt to explain the late-time accelerated expansion of the Universe.
An interesting alternative theory is modified Gauss-Bonnet gravity, or $f(G)$ gravity, where $f(G)$ is a general function of the Gauss-Bonnet term \cite{F(G)-gravity,Nojiri:2007bt}. Specific realistic models of $f(G)$ gravity were constructed to account for the late-time cosmic acceleration \cite{Nojiri:2007bt,Odintsov-models}, and it is these forms of $f(G)$ that we consider in this work. The respective constraints of the parameters of the models were also analyzed in \cite{Odintsov-models}.
It was shown that by taking into account higher-order curvature corrections the finite-time future singularities in $f(G)$-gravity are cured.
In this context, we further consider the constraints imposed by the energy conditions and verify whether the parameter range of the specific models considered in \cite{Odintsov-models} are consistent with the energy conditions \cite{Garcia:2010xz}. More specifically, we define generalized energy conditions for $f(G)$ modified theories of gravity, and consider their realization for flat Friedmann cosmological models. In particular, we analyze whether the weak energy condition is satisfied by particular choices of $f(G)$ which were advocated in Refs. \cite{Nojiri:2007bt,Odintsov-models} as leading to viable models.

\section{Gravitational field equations and the energy conditions}
\label{sec:II}
\subsection{Field equations}

The action of modified Gauss-Bonnet gravity is given by
\begin{equation}
S=\int d^4x \sqrt{-g}\left[\frac{R}{2\kappa}+f(G)\right]+S_M(g^{\mu\nu},\psi)\,,
   \label{modGBaction}
\end{equation}
where $\kappa =8\pi$. $S_M(g^{\mu\nu},\psi)$ is the matter
action, in which matter is minimally coupled to the
metric $g_{\mu\nu}$ and $\psi$ collectively denotes the matter
fields. The Gauss-Bonnet invariant is defined as
$G\equiv
R^2-4R_{\mu\nu}R^{\mu\nu}+R_{\mu\nu\alpha\beta}R^{\mu\nu\alpha\beta}$.


In this work, we consider the flat FRW background, so that the gravitational field equations for $f(G)$ modified gravity are provided by the following form
\begin{equation}
\rho_{\rm{eff}}=\frac{3}{\kappa^{2}}H^{2},
\qquad
p_{\rm{eff}}=-\frac{1}{\kappa^{2}} \left( 2\dot H+3H^{2} \right),
\label{GutenTag} \nonumber
\end{equation}
where the overdot denotes a derivative with respect $t$; $\rho_{\rm{eff}}$ and $p_{\rm{eff}}$ are
the effective energy density and pressure, respectively, defined as
\begin{eqnarray}
\rho_{\rm{eff}} &=&\rho+ \frac{1}{2\kappa^{2}} \Big[
-f(G)+24H^{2} \left(H^{2}+\dot H\right)f'(G)
    -24^{2}H^{4}
\left(2\dot H^{2}+H\ddot H+4H^{2}\dot H\right)f''(G) \Big]
 , \label{eq:rho-eff-1}\\
p_{\rm{eff}} &=&p+ \frac{1}{2\kappa^{2}}\Bigl\{f(G)-24H^{2}
\left(H^{2}+\dot H\right)f'(G)
+(24)8H^{2}\Big[ 6\dot{H}^{3}+8 H
\dot{H} \ddot{H} + 24\dot{H}^{2} H^2
   \nonumber  \\
&&+ 6H^3\ddot{H}
+ 8H^4\dot{H}+H^{2} \dot{\ddot{H}} \Big]f''(G)
 +8(24)^{2}H^{4}
\left(2\dot{H}^2+H\ddot{H}+4H^{2}\dot{H}\right)^{2}f'''(G)\Bigr\}\,.
\label{eq:p-eff-1}
\end{eqnarray}
For the flat FRW metric the curvature scalar and the Gauss-Bonnet invariant are given by $R=6(2H^{2}+\dot H )$ and $G = 24H^{2} ( H^{2}+\dot H )$, respectively.

\subsection{Energy conditions}

The energy conditions arise when one refers to the Raychaudhuri equation for the expansion. Note that the Raychaudhuri equation is a purely geometric statement, and as such it makes no reference to any gravitational field equations. The condition for attractive gravity reduces to $R_{\mu\nu}k^{\mu}k^{\nu}\geq 0$ and in general relativity, through the
Einstein field equation, one can write the above condition in terms
of the stress-energy tensor given by $T_{\mu\nu}k^\mu k^\nu \ge 0$.
Now, in $f(G)$ modified theories of gravity the field equation can be written as the following effective gravitational field equation $G_{\mu\nu}\equiv R_{\mu\nu}-\frac{1}{2}R\,g_{\mu\nu}= T^{{\rm
eff}}_{\mu\nu}$. Thus, the positivity condition, $R_{\mu\nu}k^\mu k^\nu \geq 0$, provides the following form for the null energy condition, $T^{{\rm eff}}_{\mu\nu} k^\mu k^\nu\geq 0$, through the modified gravitational field equation. As the Raychaudhuri equation holds for any geometrical theory of gravitation, we will maintain its physical motivation, namely, the focussing of geodesic congruences, along with the attractive character of gravity to deduce the energy conditions in the context of $f(G)$ modified gravity.

Thus, using the modified (effective) gravitational field equations the null energy condition (NEC), weak energy condition (WEC), strong energy condition (SEC) and the dominant energy condition (DEC) are given by
\begin{equation}
{\rm NEC} \Longleftrightarrow \rho_{\rm{eff}} +p_{\rm{eff}} \geq 0 \,,
  \label{NEC}
\end{equation}
\begin{equation}
{\rm WEC} \Longleftrightarrow \rho_{\rm{eff}} \geq 0 \; {\rm and} \; \rho_{\rm{eff}} +p_{\rm{eff}} \geq 0 \,,
   \label{WEC}
\end{equation}
\begin{equation}
{\rm SEC} \Longleftrightarrow \rho_{\rm{eff}} +3p_{\rm{eff}} \geq 0 \; {\rm and} \; \rho_{\rm{eff}} +p_{\rm{eff}} \geq 0 \,,
   \label{SEC}
\end{equation}
\begin{equation}
{\rm DEC} \Longleftrightarrow \rho_{\rm{eff}} \geq 0 \; {\rm and} \; \rho_{\rm{eff}} \pm p_{\rm{eff}} \geq 0 \,,
   \label{DEC}
\end{equation}
respectively.

In standard mechanics terminology the first four time derivatives of position are referred to as velocity, acceleration, jerk and snap. In a cosmological setting, in addition to the Hubble parameter $H=\dot{a}/a$, it is appropriate to define the deceleration, jerk, and snap parameters as
\begin{equation}
q=-\frac{1}{H^2}\frac{\ddot{a}}{a}\;, \qquad j=\frac{1}{H^3}\frac{\dot{\ddot{a}}}{a}\;,
\quad {\rm{and}} \quad s=\frac{1}{H^4}\frac{\ddot{\ddot{a}}}{a}\;.
\end{equation}

In terms of the above parameters, we consider the following definitions
\begin{eqnarray}
\dot{H}=-H^2(1+q), \qquad
\ddot{H}=H^3(j+3q+2), \qquad
\dot{\ddot{H}}=H^4(s-2j-5q-3),
\end{eqnarray}
respectively.

Using these definitions, the weak energy condition takes the following form:
\begin{eqnarray}
\rho_{\rm{eff}}+p_{\rm{eff}}&=&\rho+p+\frac{96}{k^{2}}\Big\{ -(6q^3+27q^2+21q+8qj+9j-s)f''(G)
   \nonumber  \\
&&+24[4(q^{2}+2q+1)H^{2}+2q^{2}
   +7q+j+4]f'''(G)\Big\}H^{8}\geq 0. \\
\rho_{\rm{eff}}&=&\rho+\frac{1}{2k^2}\Big[-f(G)-24H^{4} q f'(G)
-(24)^{2}H^8(2q^2+3q+j)f''(G) \Big]\geq 0.\nonumber
\end{eqnarray}

\section{Specific $f(G)$ models}

Here, we consider a realistic model of $f(G)$ gravity, which have been found to reproduce the current acceleration \cite{Nojiri:2007bt}:
\begin{eqnarray}
f(G) &=& \frac{a_1G^{n}+b_1}{a_2G^n+b_2}\,,
\label{uno}
\end{eqnarray}
where $a_1$, $a_2$, $b_1$, $b_2$ and $n$ are constants. In the following, we always assume $n>0$.

In addition to accounting for the late-time cosmic acceleration, it was also found that the specific case of Eq. (\ref{uno}) analyzed in \cite{Odintsov-models} cured the four types of finite-time future singularities emerging in the late-time accelerating era. Rather than exhaustively analyze all of the cases, we consider a specific case that does indeed prove that in addition to curing the finite-time future singularities it satisfies the weak energy condition.

For simplicity, in the example analyzed we consider vacuum, i.e., $\rho =p=0$. The WEC constraints, i.e., $\rho_{\rm{eff}}\geq 0$ and $\rho_{\rm{eff}}+p_{\rm{eff}}\geq 0$, are respectively given by
\begin{eqnarray}
&&-[a_{1}(-24qH^4)^n+b_{1}][a_{2}(-24qH^4)^n+b_{2}]+n(-24qH^{4})^n
(a_{1}b_{2}-a_{2}b_{1})
      \nonumber \\
&&\hspace{-1.2cm}+(24)^{2}nH^{8}(-24qH^{4})^{n-2}
[a_{2}(n+1)
(-24qH^{4})^{n}+b_{2}(1-n)]
\frac{(a_{1}b_{2}-a_{2}b_{1})(2q^2+3q+j)}{a_{2}(-24qH^4)^{n}+b_{2}}
\geq 0 \,, \label{NEC1a}
\end{eqnarray}
\begin{eqnarray}
&&n(-24qH^{4})^{n}(-b_{1}a_{2}
+a_{1}b_{2})\{(6q^{3}+27q^{2}+21q+8qj+9j-s)[n(a_{2}^{2}(-24qH^{4})^{2n}
-b_{2}^{2})
    \nonumber \\
&&\hspace{-0.75cm}+(a_{2}(-24qH^{4})^{n}+b_{2})^{2}]
+(4H^{2}+8H^{2}q+4H^{2}q^{2}+2q^{2}+7q+j+4)[4a_{2}b_{2}(1-n^{2})
(-24qH^{4})^{n}
    \nonumber \\
&&+a_{2}^{2}(n^{2}+3n+2)(-24qH^{4})^{2n}
+b_{2}^{2}(n^{2}-3n+2)]q^{-1}H^{-4}
\}\geq 0\,.\label{NEC1b}
\end{eqnarray}

\begin{figure}[ht]
  \centering
  \includegraphics[width=2.9in]{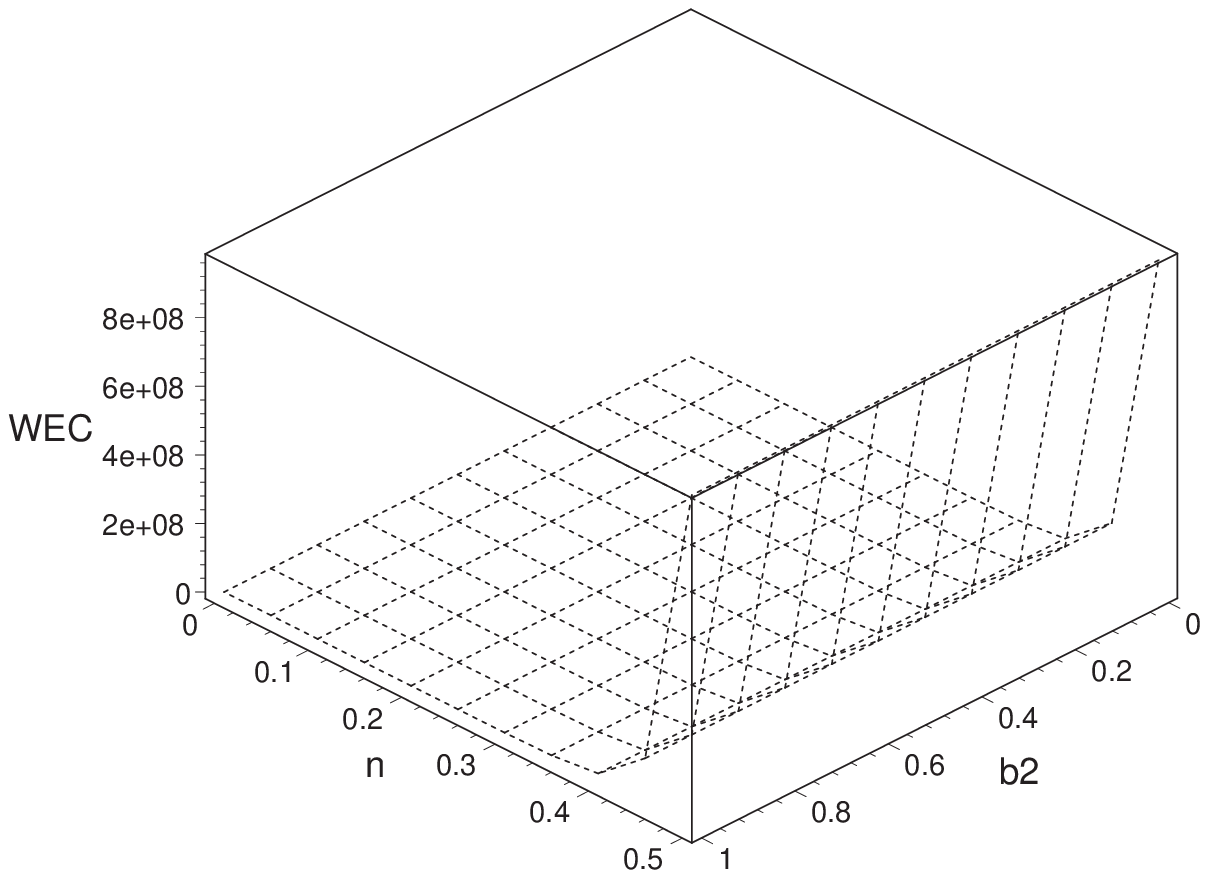}
   \includegraphics[width=2.9in]{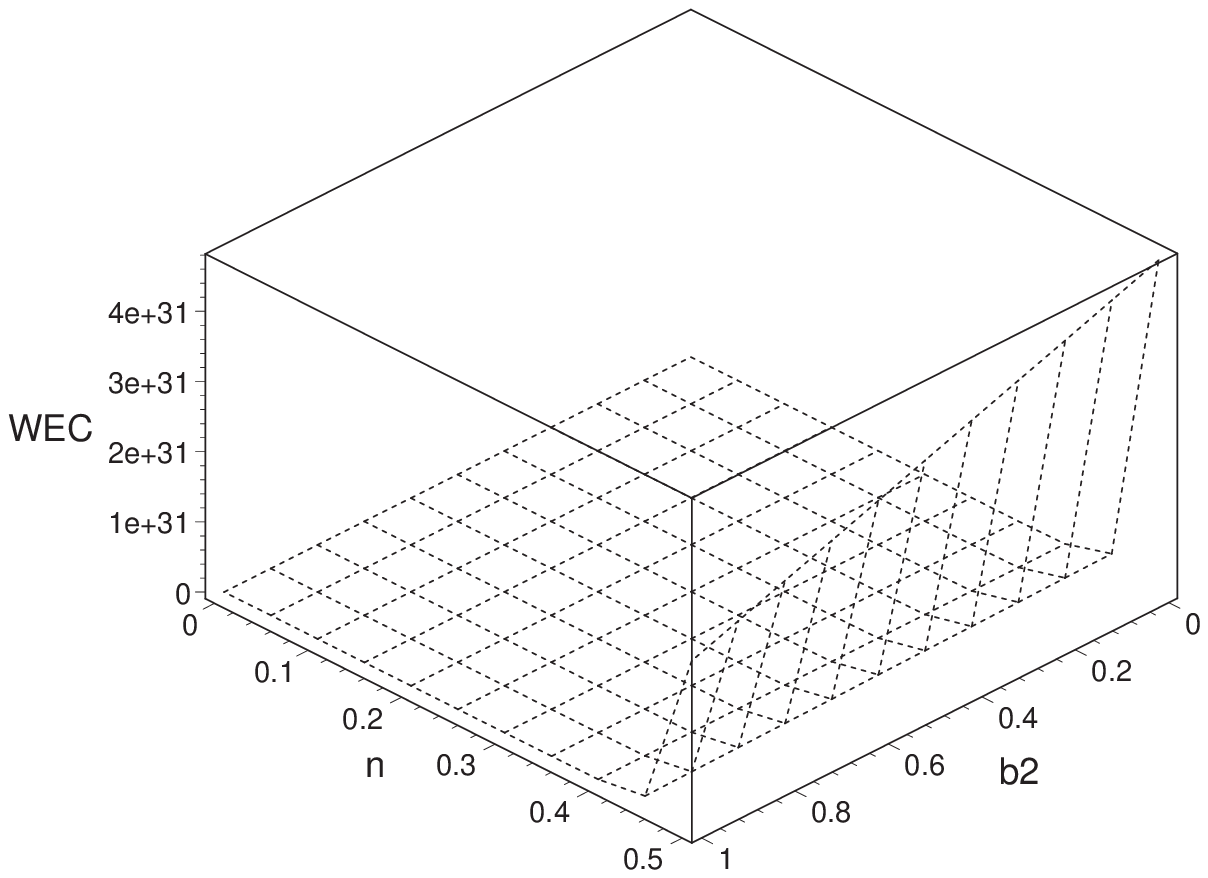}
  \caption{The plots depict the weak energy condition for the specific form of $ f(G) = \frac{a_{1}G^{n}+b_{1}}{a_{2}G^{n}+b_{2}}$. The left plot corresponds to $\rho_{\rm{eff}}\geq 0$; the right plot corresponds to $\rho_{\rm{eff}}+p_{\rm{eff}}\geq 0$. We have considered the values $a_1=-1$, $b_1=-1$, and $a_2=2$. The plots show that the weak energy condition are satisfied for the parameter range considered. See the text for details.}
  \label{fig:WEC1a}
\end{figure}


The constraints provided by the inequalities (\ref{NEC1a})-(\ref{NEC1b}) are too complicated to find exact analytical expressions for the parameter ranges of the constants $a_{1}$, $a_{2}$, $b_{1}$, $b_{2}$, and $n$, so we consider specific values for some of the parameters. In particular, we impose the following values  $a_1=-1$, $b_1=-1$, and $a_2=2$, and plot the WEC as a function of $b_2$ and $n$, which is depicted in Fig. \ref{fig:WEC1a}. The latter does indeed prove that the specific form of $f(G)$ given by Eq. (\ref{uno}) considered in \cite{Odintsov-models} is consistent with the WEC inequalities.

\section{Conclusion}

In this work, we discussed the viability of an interesting alternative gravitational theory, namely, modified Gauss-Bonnet gravity or $f(G)$ gravity. We considered a specific realistic form of $f(G)$ analyzed in the literature that accounts for the late-time cosmic acceleration and that cured the finite-time future singularities \cite{Nojiri:2007bt,Odintsov-models}. The general inequalities imposed by the energy conditions were outlined and using the recent estimated values of the Hubble, deceleration, jerk and snap parameters we have shown the viability of the above-mentioned forms of $f(G)$ imposed by the weak energy condition.


\section*{References}
\begin{thebibliography}{9}

\bibitem{expansion}
Perlmutter, S. {\it et al.} 1999, Astrophys. J.,  { 517}, 565; Riess, A. G. {\it et al.} 1998, Astron. J.,  { 116}, 1009.

\bibitem{review}
Carroll, S.M., Duvvuri, V., Trodden, M., and Turner, M.S. 2004,
    Phys. Rev.  D, { 70}, 043528;
Copeland, E.J., Sami, M and Tsujikawa, S.
  Int.\ J.\ Mod.\ Phys.\  D {\bf 15}, 1753 (2006);
%
Nojiri, S and Odintsov, S.D,
  Int.\ J.\ Geom.\ Meth.\ Mod.\ Phys.\  {\bf 4}, 115 (2007);
%
  Lobo, F.S.N.
  arXiv:0807.1640 [gr-qc];
%
De Felice, A. and Tsujikawa, S.,
  Living Rev.\ Rel.\  {\bf 13}, 3 (2010);
Nojiri, S. and Odintsov, S.D.,
  arXiv:1011.0544 [gr-qc];
%
Bertolami, O. {\it et al},
  Phys.\ Rev.\  D {\bf 75}, 104016 (2007);
  %
Boehmer, C.G., {\it et al},
  Phys.\ Rev.\  D {\bf 76}, 084005 (2007);
%
Boehmer, C.G., {\it et al},
  Astropart.\ Phys.\  {\bf 29}, 386 (2008);
%
Boehmer, C.G., {\it et al},
  JCAP {\bf 0803}, 024 (2008);
%
Bertolami, O., {\it et al},
  Phys.\ Rev.\  D {\bf 78}, 064036 (2008);
%
Harko, T. and Lobo, F.S.N.,
  arXiv:1007.4415 [gr-qc];
%
Harko, T. and Lobo, F.S.N.,
  Eur.\ Phys.\ J.\  C {\bf 70}, 373 (2010).

\bibitem{F(G)-gravity}
 Cognola, G., Gastaldi, M. and Zerbini, S.,
 Int.\ J.\ Theor.\ Phys.\  {\bf 47}, 898 (2008);
%
 Nojiri, S. and Odintsov, S.D.,
 Phys.\ Lett.\  B {\bf 631}, 1 (2005);
%
  Nojiri, S. and Odintsov, S.D.,
  Int.\ J.\ Geom.\ Meth.\ Mod.\ Phys.\  {\bf 4}, 115 (2007);
%
 Cognola, G., Elizalde, E., Nojiri, S. and Odintsov, S.D., and Zerbini, S.,
 Phys.\ Rev.\  D {\bf 73}, 084007 (2006);\
 %
 Cognola, G., Elizalde, E., Nojiri, S. and Odintsov, S.D., and Zerbini, S.,
 Phys.\ Rev.\  D {\bf 75}, 086002 (2007);\
%
 Li, B., Barrow, J.D., and Mota, D.F.,
  Phys.\ Rev.\  D {\bf 76}, 044027 (2007);\
%
 De Felice, A., and Tsujikawa, S.,
 Phys.\ Lett.\  B {\bf 675}, 1 (2009);\
%
 Alimohammadi, M., and Ghalee, A.,
 Phys.\ Rev.\  D {\bf 79}, 063006 (2009);\
%
 Boehmer, C.G., and Lobo, F.S.N.,
  Phys.\ Rev.\  D {\bf 79}, 067504 (2009);\
%
 Zhou, S.Y., Copeland, E.J., and Saffin, P.M.,
 JCAP {\bf 0907}, 009 (2009);\
%
 Goheer, N., Goswami, R., Dunsby, P.K.S., and Ananda, K.,
 Phys.\ Rev.\  D {\bf 79}, 121301 (2009);\
%
 Sadeghi, J., Setare, M.R., and Banijamali, A.,
 Phys.\ Lett.\  B {\bf 679}, 302 (2009);\
 De Felice, A., and Tsujikawa, S.,
 arXiv:0907.1830 [hep-th];\
%
Boehmer, C.G., Hollenstein, L., Lobo, F.S.N., and Seahra, S.S.,
  arXiv:1001.1266 [gr-qc].

\bibitem{Nojiri:2007bt}
 Nojiri, S., Odintsov, S.D., and Tretyakov, P.V.,
 Prog.\ Theor.\ Phys.\ Suppl.\  {\bf 172}, 81 (2008).

\bibitem{Odintsov-models}
Bamba, K., Odintsov, S.D., Sebastiani, L., and Zerbini, S.,
[arXiv:0911.4390 [hep-th]].

%
\bibitem{Garcia:2010xz}
  Garcia, N.M., Harko, T., Lobo, F.S.N., and Mimoso, J.P.,
  arXiv:1011.4159 [gr-qc].




\end{thebibliography}
\end{document}